 \documentclass[preprint,12pt]{elsarticle}

\usepackage{amssymb}

\usepackage{lineno}

\usepackage{microtype}
\usepackage{graphicx}
\usepackage{subfigure}
\usepackage{booktabs} 

\usepackage{amsmath} 
\usepackage{mathtools}

\usepackage[hyphens]{url}
\usepackage[hidelinks]{hyperref}
\hypersetup{breaklinks=true}

\graphicspath{{../Figures/}}

\journal{Journal of Marine Systems}   

\begin{document}

\begin{frontmatter}



\title{Statistical and machine learning ensemble modelling \\ 
           to forecast sea surface temperature}


\newif\ifauthordetails
\authordetailstrue

\ifauthordetails
\author[bonn]{Stefan Wolff}
\author[ibm]{Fearghal O'Donncha\corref{cor1}}
\ead{feardonn@ie.ibm.com}
\author[ibm]{Bei Chen}
\address[bonn]{PI, University of Bonn - Germany, Bonn}
\address[ibm]{IBM Research - Ireland, Dublin}
\cortext[cor1]{
Corresponding author \newline}
\fi

\begin{abstract}
In situ and remotely sensed observations have potential to facilitate data-driven predictive models for oceanography. A suite of machine learning models, including regression, decision tree and deep learning approaches were developed to estimate sea surface temperatures (SST). Training data consisted of satellite-derived SST and atmospheric data from The Weather Company. Models were evaluated in terms of accuracy and computational complexity. Predictive skill were assessed against  observations and a state-of-the-art, physics-based model from the European Centre for Medium Weather Forecasting. Results demonstrated that by combining automated feature engineering with machine-learning approaches, accuracy comparable to existing state-of-the-art can be achieved. Models captured seasonal patterns in the data and qualitatively reproduce short-term variations driven by atmospheric forcing. Further, it demonstrated that machine-learning-based approaches can be used as transportable prediction tools for ocean variables -- the data-driven nature of the approach naturally integrates with automatic deployment frameworks, where model deployments are guided by data rather than user-parametrisation and expertise. The low computational cost of inference makes the approach particularly attractive for edge-based computing where predictive models could be deployed on low-power devices in the marine environment. 
\end{abstract}


\end{frontmatter}




\section{Introduction}
Sea surface temperature (SST) is a common indicator of primary productivity in aquaculture \cite{odonncha2019precision}, critical for operation of marine-based industries such as power plants \cite{huang_coastal_2014}, while being central to better understanding interactions between the ocean and the atmosphere \cite{bayr2019error}. 
Recent decades has seen enormous progress in approaches to sample SST. In particular, satellite technology has vastly increased the granularity of measurements that are possible, providing long-term global measurements at varying spatial and temporal resolution.
MODIS (or Moderate Resolution Imaging Spectroradiometer) is a key instrument aboard the Terra and Aqua satellites, which acquire imagery data for 36 spectral bands, from which information on a range of oceanic processes, including SST, can be extracted.

Concurrently, improvements in high-resolution ocean models together with increased computational capabilities have made sophisticated data-assimilation (DA) schemes feasible -- leading to a number of reanalysis products that provide accurate forecasts across broad spatial and temporal scales. Reanalyses yield numerical estimates of the true ocean state by combining models with observations to improve short-term predictions by providing more representative initial conditions. A~state-of-the-art reanalysis is the ERA5 global dataset from the European Centre for Medium-Range Weather Forecasts (ECMWF) \cite{hirahara_era5_2016}. It provides short-term SST forecasts (and hindcasts) on a 32\,km horizontal grid at hourly intervals from a numerical synthesis of ocean models, atmospheric forcing fluxes, and SST measurements. 

These analysis and forecasting systems face a number of scientific, technical, and practical challenges. 
\begin{itemize}
\item The computational and operational requirements for ocean simulations at appropriate scales are immense and require high performance computing (HPC) facilities to provide forecasts and services in practical time frames \cite{bell_godae_2015}. 
\item Operational forecasting systems require robust data assimilation schemes that takes account of biases and errors in models and observations \citep{rawlins2007met}. 
\end{itemize} 
 A consequence of these challenges is that operational forecasting systems are only feasible for large research centres or collaborations who have access to large-scale computing resources and scientific expertise.

An alternative approach based on data-intensive computing \cite{hey2009fourth}, leverages the large datasets generated by ocean monitoring and modelling tools to train machine-learning-based forecasting models. Once trained, the computational expense of these products are negligible, and conceptually, one can develop transportable models that can be trained to learn features at different geographical location. This paper presents a suite of data-driven modelling approaches for developing robust systems to predict sea-surface temperature (SST). An automatic feature-engineering module was implemented to identify the key features at disparate geographical locations to provide a transportable forecasting system. Finally, the different models were averaged using a model-scoring and weighting approach to provide an ensemble prediction that outperformed the best-performing individual model. Contributions are as follows:
\begin{itemize}
\item We evaluated the predictive skill of a range of data-driven modelling approaches from the perspective of (1)~balancing computational complexity with predictive skill and (2)~leveraging ensemble aggregation to improve robustness.
\item We developed an autonomous feature-engineering module to (1)~improve the portability of the model to different geographical locations and (2)~reduce the appetite for training data by providing a more intelligent supply of explanatory variables.
\item Finally, we assessed performance of the modelling framework globally, against a state-of-the-art physics-based model.
\end{itemize}

While the idea of using machine learning (ML) to provide computationally cheaper surrogate models has been previously explored, the distinctive characteristics of SST lie in their complex temporal dependence structure and multi-level seasonality. To our knowledge, this application has not yet been considered in the existing literature. We demonstrate the viability of the approach to capture the short- and long-term trends: integrating different ML based models – with different temporal performance characteristics – in an ensemble approach provides accuracy on par with large scale complex models. 

In the next section, we discuss prior research in the domain. Subsequently, the different models are introduced along with the feature extraction and ensemble aggregation techniques. Section \ref{sec:results} compares performances of the different models along with the predictive accuracy of individual and ensemble aggregated models. The portability of the system to different geographical locations is discussed. Finally, we present conclusions from the research and discuss future work.

\section{Related Work}
A~wide variety of operational SST forecasting products exist that leverage physics-based circulation modelling and data assimilation to resolve temperature distributions. A~representative example is the forecasting system for the North-West Atlantic from the NEMO Community Ocean Model, which provides a variety of ocean variables at 12\,km resolution. Inputs to the system include: lateral boundary conditions from the open-ocean supplied by a (coarser) global model, atmospheric fluxes from the Met Office Unified Model and river inputs from 320 European rivers \cite{odea_operational_2012}. Other examples include the National Centers for Environmental Prediction Climate Forecast System, which provides global predictions of SST at 110\,km resolution \cite{saha_ncep_2014}, and the US Navy HYCOM Global Forecasting System, which provides 5-day forecasts at resolution ranging from 4--20\,km \cite{chassignet_godae_2009}, together with localised, regional models at higher resolution \cite{haidvogel2008ocean, chao_development_2009,odonncha2015characterizing}. The common feature of these modelling systems is the high computational demands that generally limit either the precision (coarse global models) or the size of the domain (high-resolution, regional models). 

Due to the heavy computational overhead of physical models, there is an increasing trend to apply data-driven deep-learning (DL) / machine-learning  methods to model physical phenomena \cite{bezenac_deep_2017,wiewel_latent_2018}. Application of ML-based approaches has been categorised into three areas \cite{walker_machine_2016}:
\begin{enumerate}
\item The system's deterministic model is computationally expensive and ML can be used as a code accelerator.
\item There is no deterministic model but an empirical ML-based model can be derived using  existing data.
\item Classification problems where one wish to identify specific spatial processes or events.
\end{enumerate}
A~number of studies have investigated data-driven approaches to provide computationally cheaper surrogate models, applied to such things as wave forecasting \cite{james_machine_2018}, air pollution \citep{hahnel2020using}, viscoelastic earthquake simulation \cite{devries_enabling_2017}, and water-quality investigation \cite{arandia_surrogate_2018}. 
Pertinent examples include: ML based approaches to spatially interpolate environmental variables and improve precision of solution \cite{li_application_2011}; DL-based approaches to increase the resolution of satellite imagery through down-scaling techniques \cite{Ducournau_deep_2016}; and data-mining applied to the large datasets generated by ocean monitoring and modelling tools to identify pertinent events such as harmful algal blooms \cite{Gokaraju_machine_2011}. 

Distinctive characteristics of SST are their complex temporal-dependence structure and multi-level seasonality. There are only a few options to describe systems with such characteristics, including: (1)~Generalised Additive Models (GAMs) from classic statistics, (2)~Random Forest (RF) and extreme gradient boosting (XGBoost) from ML, and (3) Multi-Layer Perceptron (MLP) and Long Short-Term Memory (LSTM) models from DL. These five models are all considered in this paper. 

\section{Machine Learning}
\label{sec:ML}
Given sufficient data, ML models have the potential to successfully detect, quantify, and predict various phenomena in the geosciences. 
While physics-based modelling involves providing a set of inputs to a model which generates the corresponding outputs based on a non-linear mapping encoded from a set of governing equations, supervised machine learning instead learns the requisite mapping by  being shown large number of corresponding inputs and outputs. In ML parlance, the model is trained by being shown a set of inputs (called features) and corresponding outputs (termed labels) from which it learns the prediction task -- in our case, given some specific atmospheric measurements we wish to predict the sea surface temperature.  With availability of sufficient data, the challenge reduces to selecting the appropriate  ML model or algorithm, and prescribing suitable model settings or \textit{hyperparameters}.  A model hyperparameter is a characteristic of a model that is external to the model and whose value cannot be estimated from data. In contrast, a parameter is an internal characteristic of the model and its value can be estimated from data during training.

Classical works in machine learning and optimisation, introduced the "no free lunch" theorem \cite{wolpert1997no}, demonstrating that no single machine learning algorithm can be universally better than any other in all domains -- in effect, one must try multiple models and find one that works best for a particular problem.
This study considers five different machine learning algorithms to predict SST. The study aims to 1) evaluate the performance of each to predict SST, 2) investigate whether simple model aggregation techniques can improve predictive skill and 3) provide insight that can be used to guide selection of appropriate model for future studies. 
While the specifics of each individual model vary, the fundamental approach consists of solving an optimisation problem on the training data until the outputs of the machine learning model consistently approximates the results of the training data. In the remainder of this section, we will describe each ML model used and provide heuristics for the selection of appropriate hyperparameters. The objective can be summarised as relating a univariate response variable $y$ to a set of explanatory variables $\textbf{x} = {x_1, x_2, ... ,x_i}$ (representing for example, air temperature, seasonal identifier, current SST, etc.).

\subsection{Generalised Additive Models}
\label{subsec:GAM}
Linear regression models are ubiquitous in statistical modelling and prediction providing a simple technique to relate predictors or features to the outcome. The relationship is linear and can be written for a single instance as: 
\begin{equation}
y = \beta_0 + \beta_1 x_1 +  ... + \beta_i x_i + \epsilon
\end{equation}
where the $\beta_i$'s are unknown parameters or coefficients that must be determined, the variables $x_i$ are features that can explain the response variable $y$, and the error $\epsilon$ is a Gaussian random variable with expectation zero.

The appeal of the linear regression model lies primarily with its simplicity and ease of interpretability. Since prediction is modelled as a weighted sum of the features, one can easily quantify the effect of changes to features on the outcome. This simplicity is also its greatest weakness since in many real-world situations: the relationship between the features and the outcome might be nonlinear, features may interact with each other, and the assumption of a Gaussian distribution of errors may be untrue.

Generalised Additive Models (GAMs) extend on linear models by instead relating the outcome to unknown smooth \textit{functions} of the features. 
 Predicting $y$ from the vector of covariates $\textbf{x}$, at time $t$ is as \cite{Hastie1990}:
\begin{equation}
g(y) = \alpha +f_1\left(x_{1}\right) + f_2\left(x_{2}\right) + ... +f_i\left(x_{i}\right) + \epsilon,
\end{equation}
where each $f_i\left(\cdot\right)$ is an unspecified function and $g(.)$  is a link function defining how the response variable relates to the linear predictor of explanatory variables (e.g. binomial, normal, Poisson) \cite{wijaya_forecasting_2015}. 

The functions $f_i\left(\cdot\right)$ can be estimated in many ways, most of which involve computer-intensive statistical methods. The basic building block of all these variations is a scatterplot smoother, which takes a scatter plot and returns a fitted function that reasonably balances smoothness of the function against fit to the data. The estimated function $f_i\left(x_i\right)$ can then reveal possible nonlinearities in the effect of the explanatory variable $x_i$. 
GAM models are particularly appealing for analysing time-series datasets in the geosciences due to interpretability, additivity of signal and regularisation: as mentioned, GAM lends itself towards interpretable models where the contribution of each explanatory variable is easily visualised and interpreted; time-series signals can be often explained by multiple additive components such as trends, seasonality and daily fluctuations which can be readily incorporated in GAM models; as opposed to simpler regression models targeted only at reducing the error, GAM admits a tuning parameter $\lambda$ that guides the "smoothness" of the model prediction (allowing us to explicitly balance the bias/variance tradeoff) \cite{friedman2001elements}. This parameter as well as the number of splines and polynomial-spline order are typically specified by the user based on heuristics, experience and model performance.

\subsection{Random Forest}
\label{subsec:RF}
Moving from statistical learning models such as GAM to those from the machine learning library, Random Forests (RF) have demonstrated excellent performance in complex prediction problems characterised by a large number of explanatory variables and nonlinear dynamics.
RF is a classification and regression method based on the \textit{aggregation} of a large number of decision trees. Decision trees are a conceptually simple yet powerful prediction tool that breaks down a dataset into smaller and smaller subsets while at the same time an associated decision tree is incrementally developed. The resulting intuitive pathway from explanatory variables to outcome serves to provide an easily interpretable model.  

In RF  \cite{breiman_randomForest_2001}, each tree is a standard Classification or Regression Tree (CART) that uses what is termed node "impurity" as a splitting criterion and selects the splitting predictor from a randomly selected subset of predictors (the subset is different at each split). Each node in the regression tree corresponds to the average of the response within the subdomains of the features corresponding to that node. The node impurity gives a measure of how badly the observations at a given node fit the model. In regression trees this is typically measured by the residual sum of squares within that node. Each tree is constructed from a bootstrap sample drawn with replacement from the original data set, and the predictions of all trees are finally aggregated through majority voting. \citep{boulesteix2012overview} 

While RF is popular for its relatively good performance with little hyperparameter tuning (i.e. works well with the default values specified in the software library), as with all machine learning models it is necessary to consider the bias-variance tradeoff -- the balance between a model that tracks the training data perfectly but does not generalise to new data and a model that is biased or incapable of learning the training data characteristics.
Some of the hyperparameters to tune include number of trees, maximum depth of each tree, number of features to consider when looking for the best split, and splitting criteria \citep{probst2019hyperparameters}. 

\subsection{XGBoost}
\label{subsec:xgboost}
 
While XGBoost shares many characteristics and advantages with RF (namely interpretability, predictive performance and simplicity), a key difference facilitating performance gain is that decision trees are built \textit{sequentially} rather than \textit{independently}. The XGBoost algorithm was developed at the University of Washington in 2016 and since its introduction has been credited with winning numerous Kaggle competitions and being used in multiple industry applications. 
XGBoost provides algorithmic improvements such as sparsity-aware algorithm for sparse data and weighted quantile sketch for approximate tree learning, together with optimisation towards distributed computing, to build a scalable tree boosting system that can process billions of examples \cite{chen_boost_2006}.

The tree ensemble model follows a similar framework to RF with prediction of the form \cite{chen_boost_2006}:
\begin{equation}
\hat{y}_i = \phi\left(x_i\right) = \sum _{k=1}^{K} f_k\left(x_i\right), \quad f_k \in \mathcal{F},
\end{equation}
where we consider $K$ trees, $\mathcal{F} = \{f\left(x\right) = w_{q\left(x\right)}\}$ represents a set of classification and regression trees (CART), $q$ represents each independent decision-tree structure, and $w_{q(x)}$ is the weight of the leaf which is assigned to the input $x$.
$\mathcal{F}$ is computed by minimising the objective function \cite{chen_boost_2006}:
\begin{equation}
\label{eqn:tree}
\begin{split}
\mathcal{L}_\phi & =  \sum_i l\left(\hat{y}_i, y_i\right) + \sum_k \Omega\left(f_k\right),  \\
& \mathrm{with} \quad \Omega\left(f\right) =  \frac{1}{2} \lambda \|w\|^2,
\end{split}
\end{equation}
with $l$ being a differentiable convex loss function (for example the mean squared error) of the difference between the prediction  $\hat{y}_i$ and the observation $y_i$ for each realisation $i$. The regularisation term, $\Omega$, smooths the final weights to avoid over-fitting ($\lambda$ is a regularisation coefficient). Furthermore, a restriction to a maximal tree depth serves to regulate model complexity.

\subsection{Multi-Layer Perceptron}
\label{subsec:MLP}

The first DL-based approach investigated was a Multi-Layer Perceptron (MLP) model. An MLP network solves an optimisation problem to compute the weights and biases that represent the nonlinear function mapping inputs to the best representation of outputs, $\mathbf{\hat{{{y}}}}$:
\begin{equation}
\label{eqn:ANN}
	  	g\left(\mathbf{x};\mathbf{\Theta}\right)=\mathbf{\hat{y}}.
\end{equation}
$\mathbf{\Theta}$ denotes the mapping matrix of weights and biases that represents the relationship between SST and explanatory variables, $\mathbf{x}$ in the form of a neural network. 

\begin{figure}[ht!]
\centering
\includegraphics[width=1.0\textwidth]{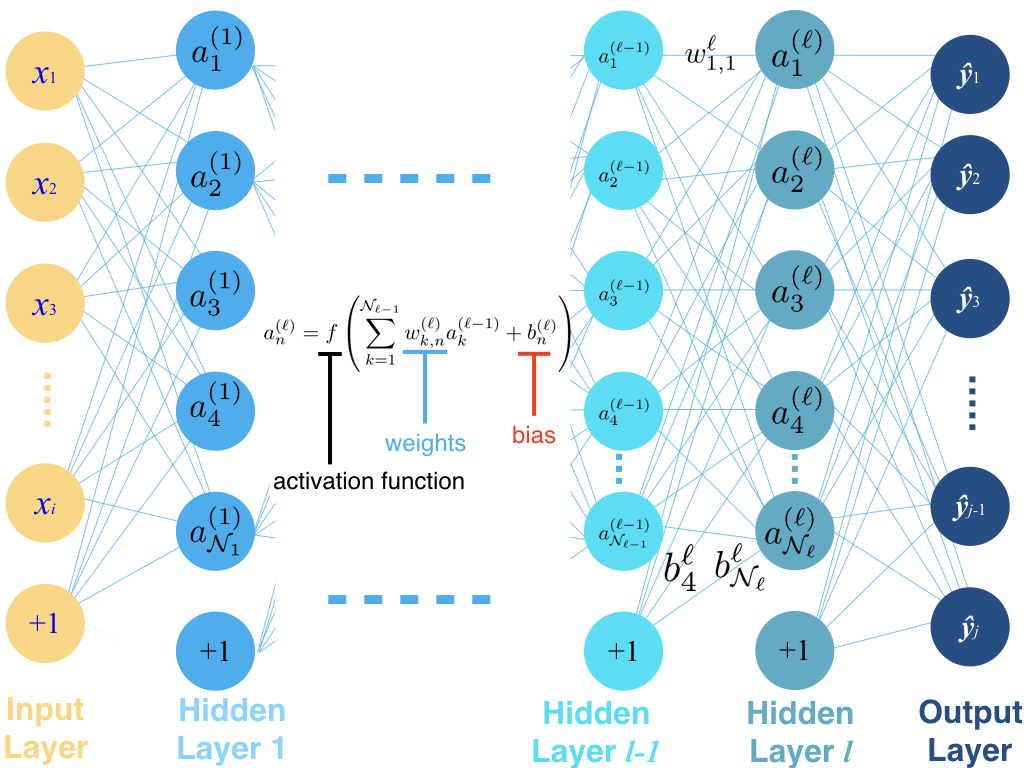}
\caption{Schematic of an MLP machine learning network as illustrated in \citep{james_machine_2018}.}
\label{fig:deeplearning}
\end{figure}

An MLP model is organised in sequential layers made up of interconnected neurons. As illustrated in Figure~\ref{fig:deeplearning}, the value of neuron $n$ in hidden layer $\ell$ is calculated as:
\begin{linenomath*}
\begin{equation}
a_{n}^{\left(\ell\right)} = f\left(\sum^{\mathcal{N}_{\ell-1}}_{k=1} w_{k,n}^{\left(\ell\right)} a_{k}^{\left(\ell-1\right)} + b_{n}^{\left(\ell\right)}\right),
\end{equation}
\end{linenomath*}
\noindent where $f$ is the activation function, $\mathcal{N}_{\ell-1}$ is the number of nodes in layer $\ell-1$, $w_{k,n}^{\left(\ell\right)}$ is the weight projecting from node $k$ in layer $\ell-1$ to node $n$ in layer $\ell$, $a_{k}^{\left(\ell-1\right)}$ is the activation of neuron $k$ in hidden layer $\ell-1$, and $b_{n}^{\left(\ell\right)}$ is the bias added to hidden layer $\ell$ contributing to the subsequent layer.  The activation function selected for this application was the rectified linear unit (ReLU) \citep{nair2010rectified}:
\begin{linenomath*}
\begin{equation}
f\left(z\right)=\max\left(0,z\right).
\end{equation}
\end{linenomath*}

A~loss function is defined in terms of the squared error between the observations and the machine-learning prediction plus a regularisation contribution controlled by $\lambda$:
\begin{linenomath*}
\begin{equation}
\label{eqn:lossfunction}
\vartheta=\frac{1}{2}\sum_{k=1}^m\lVert{{\mathbf y}}^{\left( k\right)}-{\hat{{\mathbf{y}}}}^{\left( k\right)}\rVert_2^2 + \lambda \lVert{{\boldsymbol{\Theta}}}\rVert_2^2,
\end{equation}
\end{linenomath*}
where the $||\cdot||_2$ indicates the $L_2$ norm. The regularisation term  penalises complex models by enforcing weight decay, which prevents the magnitude of the weight vector from growing too large because large weights can lead to overfitting -- a condition where the model fits the training data well but does not generalise to new data \citep{goodfellow2016deep}.

By minimising the loss function, the supervised machine learning algorithm identifies the ${{\boldsymbol{\Theta}}}$ that yields $\hat{{\mathbf{y}}}\approx{\mathbf y}$. As shown in Figure~\ref{fig:deeplearning}, a machine learning algorithm transforms an input vector (layer) to an output layer through a number of hidden layers. The machine learning model is trained on a data set to establish the weights parameterising the space of nonlinear functions mapping from ${\mathbf x}$ to ${\mathbf y}$.

Hyperparameter tuning is required to balance the effective capacity of the model and the complexity of the task. In neural network type approaches, increasing the number of layers and of hidden units per layer increases the capacity of the model to represent complicated functions. Hence increasing the depth of the network can improve performance on the \textit{training} data but run the risk of overfitting --  thereby reducing generalisation potential. Standard hyperparameters to tune in neural networks include the number of layers, number of nodes and the regularisation coefficient $\lambda$.

\subsection{Long short-term Memory Model}
\label{subsec:lstm}

Cognisant of the temporal nature of the data, we investigated the performance of recurrent neural network (RNN) type models. A~fundamental extension of RNNs compared to MLP is parameter sharing across different parts of the model. This has intuitive applicability to the forecasting of time-series variables with historical dependency. An RNN with a single cell recursively computes the hidden vector sequence $\mathbf{h}$ and output vector sequence $\mathbf{y}$ iteratively from $t=1,\ldots,T$ in the form \cite{Graves2013}:
\begin{equation}
\begin{split}
 & h_t = \mathcal{H} \left(W_{xh}x_t + W_{hh}h_{t-1} + b_h \right), \\
 & y_t = W_{hy}y_t +b_y.
\end{split}
\end{equation}
where the $\mathbf{W}$ terms denote weight matrices (e.g. $W_{xh}$ is the
input-hidden layer weight matrix), the $\mathbf{b}$ terms denote bias vectors (e.g. $\mathbf{b}_h$ is hidden layer bias vector) and $\mathcal{H}$ is the hidden layer function which is typically implemented as a sigmoid function. In effect, the RNN has two inputs, the present state and the past.

Standard RNN approaches have been shown to fail when lags between response and explanatory variables exceed 5--10 discrete timesteps \cite{gers_lstm_1999}. Repeated applications of the same parameters can give rise to vanishing, or exploding gradients leading to model stagnation or instability \cite{goodfellow2016deep}.
A number of approaches have been proposed in the literature to address this, with the most popular being LSTM.

Instead of a simple weighted dependency, `LSTM cells' also have an internal recurrence (a self-loop), that serves to guide the flow of information and reduce susceptibility to vanishing or exploding gradients. 
Each cell has the same inputs and outputs as an ordinary recurrent network, but also has more parameters and a system of gating units that controls the flow of information. An LSTM model has a number of gates: input, output and forget gates that decide whether to let information in, forget information because it is not important, or let it impact output at the current timestep, respectively. As new input comes in, it's impact can be accumulated to the cell, forgotten or propagated to the final state depending on the activation of the relevant gates \cite{Shi_2015_CNNLSTM}. In analogy to the MLP, we use L2-regularisation of weights represented by the parameter $\lambda$, in an equivalent manner to equation \ref{eqn:lossfunction}. More details on LSTM are provided in  \citet{gers_lstm_1999}.

\subsection{Feature Engineering}
\label{sec:feature_engineering}
In traditional modelling based on solving a set of partial differential equations (PDE), the relationship between inputs and outputs are clear -- founded on well-understood physics. Machine learning on the other hand relies on the concept of learning complex, nonlinear relationships between inputs and outputs. While the outputs are clear (the variable we wish to predict), the inputs are more opaque and one wishes to consider all variables that potentially contribute to the output response, while avoiding superfluous data that may hinder performance. When predicting SST, some of the variables that may contribute include a wide range of atmospheric conditions (air temperature, solar radiation, cloud cover, precipitation, wind speed, etc.), autoregressive features (i.e. past values of the response variable -- SST), temporal information (e.g. season, day of year, time of day), and potentially values at neighbouring spatial locations.  Feature engineering is the process of using domain expertise and statistical analysis to extract the most appropriate set of features for a particular problem from the entire set of data that may contribute. The role of feature engineering is to improve predictive accuracy and expedite model convergence by selecting the most appropriate features that explain the response variable and provide maximum value. Excluding important data will limit the predictive skill of the model while superfluous data tends to add noise to the model. 

\begin{figure}[h!]
\centering
\includegraphics[width=0.8\textwidth]{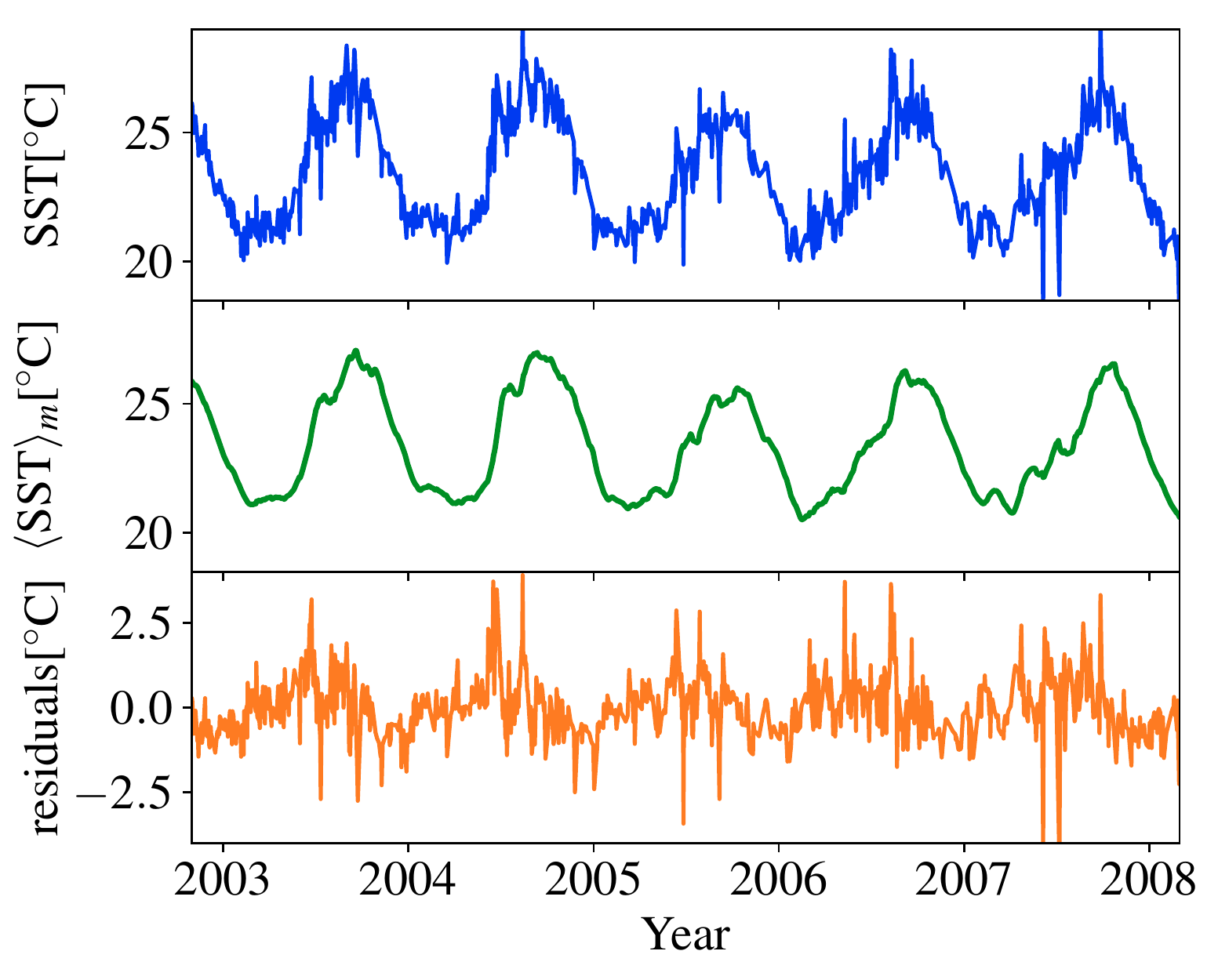}
\caption{SST time series from MODIS measurements (upper panel), monthly rolling mean (middle panel), and residuals after subtraction of the monthly rolling mean from the SST data (lower panel).}
\label{fig:data_exploration}
\end{figure}

Figure \ref{fig:data_exploration} shows multi-year SST data illustrating primary patterns. A~monthly rolling mean of the data (middle plot) was subtracted from the raw data (top plot) with residuals presented (bottom plot). The seasonal pattern of the data is evident with yearly cycle capturing a significant portion of the data variance. The data residuals largely represent short-term fluctuations in the data (together with sensor uncertainty component). The objective of the modelling was to learn the nonlinear relationships between the explanatory variables and the long- and short-term signals of the data.

For machine learning forecasts, the raw data themselves are rarely the most informative and a number of combinations and transformations of the raw data must be considered. The feature variables used for this study consisted of SST historical time series data from MODIS Aqua satellite,  atmospheric data from The Weather Company (TWC), and time features (season, day of year, etc.). 
From these raw data, several different types of features were designed and investigated. The feature engineering process combined domain expertise to initially select known variables influencing SST, with statistical analysis to explore strength of relationship between a large number of features and the response variable. The dependence or correlation between each feature and response variable was determined based on a univariate feature selection using the SciKit-Learn \citep{sklearn_api} feature selection library. A subset of features with highest $F$-scores \cite{guyon2008feature} were retained. The implementation of the feature selection approach is described in more detail in Section \ref{subsec:modsetup}.

\section{Methodology}
\label{sec:methodology}

Application of ML techniques can be reduced to a number of steps related to: selection of appropriate ML algorithms, providing sufficient data to train the models, and selecting the correct model hyperparameters (settings or parameters that must be defined outside the learning algorithm) for the model. In the remainder of this section, we will describe the training data used, provide details on each model considered and outline the application of each model to the problem of forecasting SST.

\subsection{Input Data}
\label{sec:data}

Training data were from the MODIS instrument aboard the NASA \textit{Aqua} satellite. MODIS SSTs are produced and made available to the public by the NASA GFSC Ocean Biology Processing Group. The MODIS sensor measures ocean temperature (along with other ocean products such as salinity and Chlorophyll concentration) from a layer less than 1\,mm thick at the sea surface. Data are available from 2002 to present at 4\,km horizontal resolution and daily intervals \citep{obpg_modis}. Calibration of the Pathfinder algorithm coefficients and tuning of instrument configurations produce accurate measurements of SST with mean squared error (MSE) against \textit{in situ} sensors $<\mathrm{0.2}^\circ$C \cite{kilpatric_modis_2015}. These accurate global SST measurement over a multi-decade period, serve as an ideal dataset to extract insights using ML. Daily, weekly (8 day), monthly and annual MODIS SST products are available at both 4.63 and 9.26 km spatial resolution and for both daytime and nighttime passes. The particular dataset we used was the MODIS Aqua, thermal-IR SST level 3, 4km, daily, daytime product downloaded from the Physical Oceanography Distributed Active Archive Center (PODAAC)  \citep{obpg_modis}. This MODIS SST data served as \textit{labels} to the machine learning algorithm while the data were also used as autoregressive (lagged) \textit{features} to the model.

As described in section \ref{sec:feature_engineering}, various combinations of atmospheric variables were provided as features to the model, extracted from The Weather Company through their public API \cite{IBM2018}. The variables used were the 18 atmospheric quantities included as part of the standard weather variables described in the API documentation \citep{TWC-cleanedHistorical}.
While we do not have rights to redistribute The Weather Company data, 
a free API key can be obtained to download the data from the vendor.

A key part of any modelling study is validation of the prediction and comparison against benchmark values. While not provided as inputs to the models, we used data from ECMWF model data to assess predictive skill. ECMWF ERA5 dataset provides an atmospheric reanalysis of the global climate at 32\,km horizontal grid at hourly intervals from a numerical synthesis of ocean models and atmospheric forcing fluxes \cite{hirahara_era5_2016}.  We downloaded SST data at the nearest grid cell to the MODIS dataset using the ECMWF Climate Data Store API (CDSAPI) to serve as a validation dataset.

\subsection{Model Setup and Training}
\label{subsec:modsetup}
As described in Section \ref{sec:ML}, there are three primary steps to deployment of a machine learning model:
\begin{itemize}
\item Feature engineering, where the requisite explanatory data are extracted, processed and combined to be fed to the model (described in Section \ref{sec:feature_engineering}).
\item Selection of the most appropriate hyperparameters for the model.
\item Training the model by feeding the training data to the model which finds patterns in the data that map the input data attributes to the target.
\end{itemize}

The first two steps were conducted on a dataset extracted from an arbitrary location in the North Atlantic: ($27^\circ 28^\prime 46.45^{\prime\prime}$ N, $32^\circ 25^\prime43.71^{\prime\prime}$ W). 
MODIS SST data were collected over 16 years from July 2002 (earliest available data) to December 2018. 
Satellite measurements are prone to missing data --  for instance due to cloud cover. For this location, data was missing 57\% of days, which was representative of data availability at other locations also. As data gaps are problematic for the training of time-series models (due to auto regressive dependencies), linear interpolation between adjacent values replaced the missing data. While this can introduce artefacts to the data, the fact that missing values were evenly distributed across the entire dataset and the time series nature of the data made interpolation the best approach. Moreover, the secondary weather input, TWC reanalysis data, were complete. 

As previously described, the experiments compared a number of feature selection approaches to incorporate atmospheric and autoregressive effects. Initially the most appropriate number of lags to specify as  autoregressive SST features were  selected based on heuristics (different temporal scales involved such as daily, seasonal, year) and a trial-and-error search of a limited number of possible lags. To simplify analysis of lag selection, the models were considered as autoregressive models for this stage (i.e. we only supplied SST at previous timesteps as inputs and did not include weather data). We observed that these simple autoregressive models provided adequate predictive skill for short-term forecasting of up to two days (for longer-term predictions, atmospheric features were critical for performance). Nevertheless, this simplified modelling study enabled insight into the most suitable number of lags (or number of AR steps) to include in each model deployment. For the GAM, RF and XGBoost models, the optimal lags were found to be approximately 30 days, which balanced computational tractability with predictive skill. To incorporate seasonal effects (and also due to greater computational efficiency), the MLP and LSTM model were fed data from up to the previous 400 days (to extend beyond one year of historical trend). It's worth noting that when including AR features, it introduces a temporal dependency which is important if one wishes to make forecast multiple days in advance -- i.e. to make forecast for day $t+2$, predicted SST for day $t+1$ is provided as a feature. This allowed for long-term prediction but introduced the possibility of systematic model error and bias (i.\,e.~prediction error accumulated). This is analogous to model drift observed in numerical modelling studies where model forecast can diverge from true state over time \citep{doi:10.1175/1520-0469(1963)020<0130:DNF>2.0.CO;2}.

The AR features described above were combined with time features (season, month and week of year), and various combinations of atmospheric features to construct different model scenario inputs.  Weather feature were selected from atmospheric data consisting of 18 time-dependent atmospheric quantities reporting standard meteorological variables such as, air temperature, solar radiation flux, cloud cover and winds  \cite{IBM2018}. Three different model scenarios were created from these with different combinations of atmospheric data, namely:
\begin{itemize}
\item all 18 atmospheric quantities at the desired time were fed to the model (we refer to this scenario as TWC1).
\item To reduce the number of covariates (and hence network size and associated demands for training data), a feature-selection module quantified the most important variables. Univariate feature selection was performed by computing 
$F$-scores from the correlation of each single features with the output label \cite{guyon2008feature} and retaining the \textit{three}  atmospheric features with highest scores as described in Section \ref{sec:feature_engineering} (referred to as TWC2).
\item This concept was further extended with time-dependent information by assigning univariate scores to lagged values of the selected atmospheric features in scenario 2, and choosing the lags with the highest scores as features (referred to as TWC3). This reflected that SST is also likely to be influenced by atmospheric conditions (e.g. air temperature) at previous days.
\end{itemize}
The resultant set of features to be considered for each model were, AR features with specific lag (AR), time features (time) and most appropriate combination of weather features (TWC1, TWC2 and TWC3). For all five models these  set of features were investigated and for each model the best performing combination were selected that minimised error against the test dataset. Emanating from the different characteristics and complexities of the models it was not expected that a single feature combination would provide best performance across all models. Instead effective machine learning implementations requires a careful balance of appropriate features, model or algorithm complexity, and hyperparameter selection. 

The next stage of model setup considered hyperparameter optimisation for each of the models. In general machine learning models have a number of hyperparameters, and the selection of the most appropriate is a combination of heuristics, expertise and trial-and-error. For each model, hyperparameter optimisations adopted a greedy, grid-search approach over the user-defined parameter ranges summarised in Table~\ref{tab:hyper_params}. The $\mathbf{x}$ and $\mathbf{y}$ data were split into two groups, to form the training-data set composed of 90\% of the 6018 rows of data, and the test-data set the remaining 10\%. For each model the learning algorithm was trained on the training data and then applied to the test data set and the MSE between test data vector, $\mathbf{y}$, and its machine-learning representation $\mathbf{\hat{y}}$ was calculated. The hyperparameter combination that minimised this MSE was selected for each model.  The selected values are presented in Table \ref{table:hp_selected} and discussed in more detail in Section \ref{sec:results} where we evaluate model performance.

\begin{table}[t!]
  \caption{Hyperparameters and ranges used for model design. See Section \ref{sec:ML} for details on each model hyperparameter} 
  \centering
  \begin{tabular}{ll}
  \toprule
	\textbf{Model} & \textbf{Hyperparameters} \\ \midrule
	GAM & \# of splines/features $\in \{10, 15, 20\}$ \\
	    &polynomial-spline order $\in \{3, 5, 8\}$ \\
	 & $\lambda \in \{0.001, 0.01, 0.1, 1, 10, 100\}$ \\ \rule{0pt}{3ex}

	RF & \# of trees/features $\in \{100, 200, 500\}$ \\
	    & max \# of features $\in \{3, 5, 10\}$ \\ 
	 &  max depth $\in \{5, 10, 15, 20 \}$ \\ \rule{0pt}{3ex}

	XGBoost & \# of trees/features $\in \{500, 700, 1000\}$ \\
	 &  max depth $\in \{5, 10, 15, 20 \}$ \\
	 &  $\lambda \in \{ 0.01, 0.05, 0.1, 0.5\}$ \\ \rule{0pt}{3ex}

	MLP & \# of layers $\in \{ 5, 10, 20\}$ \\
	 & \# of nodes/layer $\in \{ 20, 50, 75 \}$ \\
	 &  $\lambda \in \{ 0.001, 0.01, 0.1, 1\}$ \\ \rule{0pt}{3ex}

	LSTM & \# of layers $\in \{ 1, 2\}$ \\
	 & \# of units/layer $\in \{ 1, 2, 3 \}$ \\
	 &  $\lambda \in \{ 0.001, 0.01, 0.1, 1\}$\\
 \bottomrule
  \end{tabular}
\label{tab:hyper_params}
\end{table}

A number of Python toolkit libraries were used to access high-level programming interfaces to statistical and machine learning libraries and to cross validate results. The GAM model was implemented using the LinearGAM API from pyGAM \citep{daniel_serven_2018}, Random Forest from the widely-used SciKit-Learn \citep{sklearn_api} toolkit, and XGBoost from the python implementation of the software library \citep{chen_boost_2006}. The deep learning models, namely MLP and LSTM were implemented using the popular Keras library which serves as a high-level neural network API \citep{chollet2015keras}.

\subsection{Model Scoring and Aggregation}
To assess the different modelling approaches, hyperparameters, and combinations of input features, the time series was split into training (90\%) and test (10\%) sets.
The models were trained to make a prediction one day ahead based on feeding the previously described features and labels (measured value of SST for that day). The test datasets were then used to evaluate the performance of the model prediction against measured values.

 As prediction depended on historic estimates of SST (i.e. a prediction for one day ahead required information on the current SST), the model prediction was fed back as a feature to the model in a recurrent fashion. Specifically, the first test prediction ($t=1$) was made with \textit{measured} values of SST (at time $t=0$) as a feature. For future predictions, the measured SST feature was replaced with the \textit{prediction} from the previous day, i.e. prediction at time $t=2$ received as input feature, model prediction for time $t=1$ instead of the satellite derived value of SST, which in practise would not be available for forecasting multiple days in advance. Scoring for the entire test dataset (562 days) proceeded in this manner. This made the study sensitive to propagation of error, where a low skill prediction propagates through the entire forecasting period (in a similar manner that error in initial condition or boundary forcing formulation can propagate through the prediction of a physics-based model).

 The mean average error (MAE) and mean absolute percentage error (MAPE) assessed the accuracy of each model:
 \begin{align}
\textsc{MAE} = \frac{1} {N_\mathrm{test} }  \sum_{i=1}^{N_\mathrm{test}} \left\vert  \left(y - \hat y\right)  \right\vert,
&&
  \textsc{MAPE} = \frac{100}{N_\mathrm{test}}\sum_{i=1}^{N_\mathrm{test}} \left\vert \frac{\left(y - \hat y\right)  }{y} \right\vert, 
\end{align}
\noindent
where $N_\mathrm{test}$ is the size of the training data, $y$ is the measured data, and $\hat{y}$ the model-predicted equivalent. 
Finally, the models were aggregated into a single best prediction weighted by the inverse MAPE of the test data \cite{Adhikari2012_inverseMAPE_average}. Convex weights for the models considered in this study were computed of the form:
\begin{equation}
\label{eqn:invMape}
W_m = \left(\frac{MAPE_m}{\sum_m^p(MAPE_m)}\right)^{-1}
\end{equation}
where $W_m$ is the weight for model $m$, $p$ is the number of models, and $MAPE_m$ is the MAPE of model $m$.

\section{Results}
\label{sec:results}

For each model implementation, we focused on identifying the optimal combination of features and hyperparameters that maximise predictive skill. 
Table~\ref{tab:model_metrics} presents model-selection results considering hyperparameters and feature engineering. The combination of model complexity and size of the features datasets are evident. Relatively simple models like GAM and RF provided best performance with more sophisticated feature engineering that reduced the size of the dataset. However, MLP and XGBoost both yielded the lowest test MAPE when provided with the full atmospheric dataset and allowed to infer relationships from all variables and data labels. 


In addition to MAPE accuracy, Table~\ref{tab:model_metrics} also lists the run times needed to train the corresponding models (on a commodity laptop). Training times were within acceptable limits for all models, although significant variability existed. As expected, the LSTM had the largest computational demand -- however it also had the highest MAPE. This non-intuitive result demonstrates the need to balance model complexity with the nature of the data. That is, na\"{i}ve selection might suggest that an RNN-based model such as LSTM is most suitable for a time-series dataset. However, results demonstrated that the LSTM model failed to capture the high-frequency variations in the data and only captured the general seasonal patterns (the monthly rolling-mean trends reported in Figure \ref{fig:data_exploration}). The inability to capture short-scale variations is due to the ``long memory'' for this model that interfered with learning short-term variations. In contrast, simpler models with time-series information explicitly included as features better learned short-term dynamics. 

\begin{table}[t]
\label{table:hp_selected}
  \caption{Model-selection result for the two North Atlantic locations.}
  \centering
  \resizebox{\textwidth}{!}{%
  \begin{tabular}{lllcc}
	\toprule
	\bf{Model} & \bf{Features} & \bf{Hyperparameters} & \bf{Training time [sec]} & \bf{MAPE}\\ \midrule
	\small
	GAM & TWC3, time &  \# of splines = 20, spline order = 8, $\lambda =10$ & ~~1.67 & 2.27 \\
	RF & AR, TWC3, time &  \# of trees = 500, max \# of features = 3, max depth = 20 & 19.09 & 1.97 \\
	XGBoost & TWC1, time &  \# of trees = 1,000,  max depth = 5, $\lambda=0.05$ & ~~0.21 & 2.17 \\
	MLP & TWC1, time & \# of layers = 20, of units/layer = 20, $\lambda=0.01$ & 11.18 & 2.07 \\
	LSTM & TWC2 & \# of layers = 1, \# of units/layer = 2, $\lambda=0.1$ & 71.67 & 2.54\\
 \bottomrule
  \end{tabular}
  }
  \label{tab:model_performance}
\label{tab:model_metrics}
\end{table}

\begin{figure}[h!]
\centering
\includegraphics[width=0.7\textwidth]{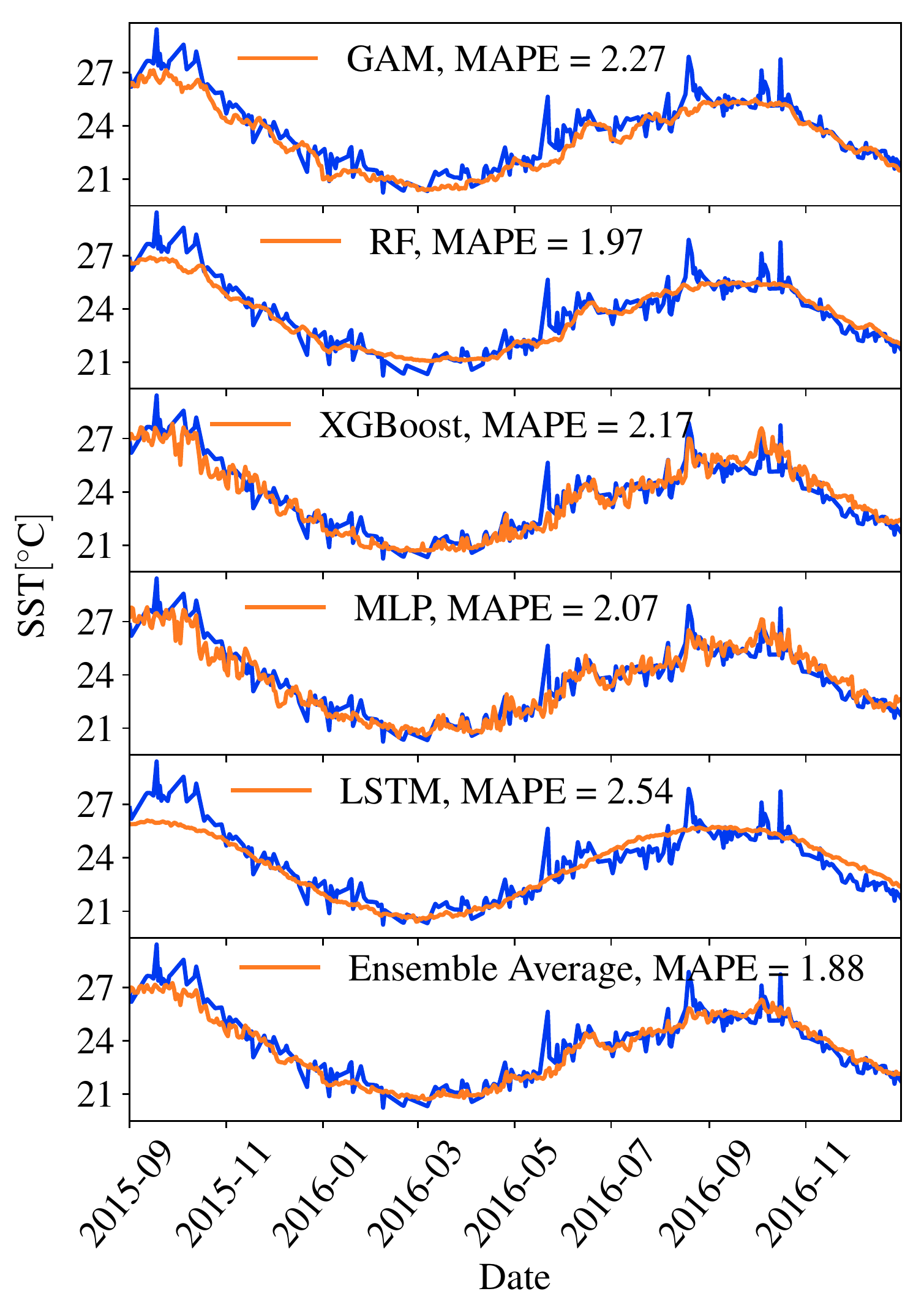}
\caption{Test-prediction (orange curve) of SST at $27^\circ 28^\prime 46.45^{\prime\prime}$ N, $-32^\circ 25^\prime 43.71^{\prime\prime}$ W from different ML models trained on 12 years of preceding historical data compared to measured SSTs (blue curve). Bottom figure presents ensemble average of all models based on individual model MAPE. Feature combinations and hyperparameters adopted for each model are summarised in Table \ref{tab:model_metrics}.}
\label{fig:different_models}
\end{figure}

Figure~\ref{fig:different_models} compares model predictions (see Table~\ref{tab:model_metrics}) for the test period to MODIS data. Observing the time evolution of SST reveals that a suitable model must represent two distinct time scale components. On the one hand there is the smooth SST evolution governed by seasonality. This component of SST evolution benefited from suppression of large fluctuations. Of the models studied, this criterion was fulfilled by the GAM approach, which yielded lowest test MAPE with a large regularisation parameter $\lambda = 10$ (i.\,e.,~the penalty on the second-order derivative of fitted single-feature functions). 
The large regularisation effect together with the piecewise polynomial components of GAM models contributed to a smoother time-series prediction that still captured long-term trends, including correlation of data between years. Similarly, the RF approach led to a comparably smooth SST evolution but at significantly lower MAPE than the GAM model.
The most obvious reflection of the seasonal pattern is evident in the LSTM prediction which produces a highly smoothed representation of the training data. 
The model fails to capture any small-scale dynamics at the daily or weekly level instead reproducing the seasonal heating/cooling effects only. Further analysis of model parameters suggested this to be a result of the retained long-term memory informing the broader trend only.  On the other hand, the seasonal cycle has superimposed on it short-term behaviour dominated by peak events occurring at daily to weekly time scales. This is particularly evident in the XGBoost and the MLP approaches where both yielded best performances for smaller $\lambda$, which enabled them to better capture short-term events. It's worth noting that while XGBoost and MLP captured the small-scale fluctuations better, RF returned lowest MAPE. To simultaneously take both aspects into account, the final plot aggregates models based on an inverse MAPE weighting  as presented in equation \ref{eqn:invMape}. As a preprocessing step, due to the comparably poor performance of the LSTM model, it was excluded from the ensemble.  The ensemble average generates lowest MAPE indicating that a relatively simple model-weighting aggregation approach can outperform an individual best performing model  \citep{odonncha2018integrated, o2019ensemble}

\begin{figure}[h!]
\centering
\includegraphics[width=0.9\textwidth]{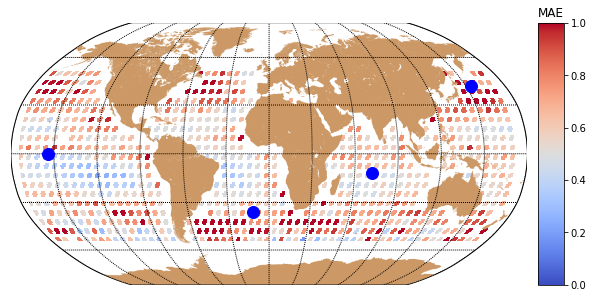}
\includegraphics[width=0.9\textwidth]{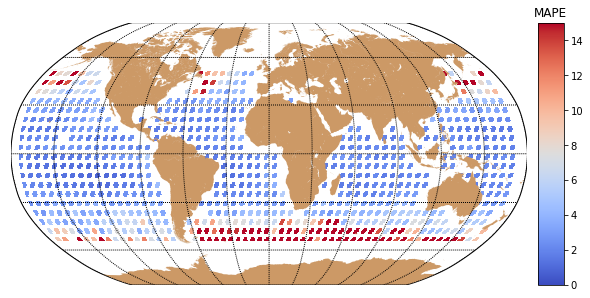}
\caption{Predictive skill of a weighted ensemble average of GAM, RF, XGBoost, and MLP models at a set of locations distributed equally between $\pm$ 54$^\circ$ latitude. MAE (top) and MAPE (bottom) metrics are presented to inform on absolute and relative errors. The features and hyperparameters prescribed are presented in Table~\ref{tab:model_metrics}.
The blue circles on the top plot denote locations that are analysed in more detail in Section \ref{sec:r&d} and presented in Figure \ref{fig:simulation_comparison}.  }
\label{fig:globalskill}
\end{figure}

\subsection{Transportability and Comparison to State-of-the-art}
As the feature engineering and hyperparameter selection process is complex and cumbersome, it is desirable to execute this procedure once and then use the selected model at different locations. The objective being to identify the most appropriate model inputs (features) and settings (hyperparameters) from a small dataset, which are then used to train (on new data) and deploy (i.e. make forecast) the models at any location we wish to make SST forecast. 
We investigated the performance of the model at a set of globally distributed locations. 
Data (SST measurements and TWC weather variables) were collected in a 6$^\circ \times$ 6$^\circ$ grid of points within 1$^\circ$ of shorelines between $\pm$ 54$^\circ$ latitude. This resulted in 730 locations globally. 
While the features and hyperparameters were selected as noted in Table~\ref{tab:model_performance}, the models were retrained at each location in a similar manner as previously described using a 90\%/10\% train and test data split.  The resulting prediction was again an aggregation of GAM, RF, XGBoost, and MLP results, where each model was weighted by the inverse MAPE at each location to favour models with better performance in the weighted average. Figure~\ref{fig:globalskill} presents the MAE and MAPE computed at these 730 locations. 
Results demonstrate that MAE and MAPE were less than 1$^\circ$C and 10\%, respectively, at most locations. Table \ref{tab:globmape} presents average error metrics over all locations. The MAPE-weighted ensemble average returned MAE and MAPE of 0.68$^\circ$C and 7.9\% respectively. These values are comparable to ECMWF estimates on SST which returned values of 0.56$^\circ$C and 12.3\%, respectively. 
While ECMWF reports lower absolute error, relative errors are noticeably higher. This suggests a tendency of the numerical outputs to perform poorer in periods when temperatures are lower (increasing relative error). 

\begin{table}[h!]
  \caption{MAE and MAPE averaged across all spatial locations presented in Figure~\ref{fig:globalskill}. Metrics are presented for each model individually, an ensemble averaged weighted by the inverse MAPE and the final column presents error metrics for  a benchmark ECMWF model against MODIS measurements.}
  \centering
  \begin{tabular}{c c c c c c c}
  \toprule
	\textbf{Metric} & \textbf{GAM} & \textbf{RF} & \textbf{XGBoost} & \textbf{MLP} & \textbf{Ens. Ave.} &  \textbf{ECMWF} \\ \midrule
	MAE & 0.78  & 0.72 & 0.79 & 0.89 & 0.68 & 0.56  \\ 
    MAPE & 9.7  & 9.4 & 8.8 & 10.6 & 7.9 & 12.3\\   
\bottomrule
  \end{tabular}
\label{tab:globmape}
\end{table}

\begin{figure}[h!]
\centering
\includegraphics[width=\textwidth]{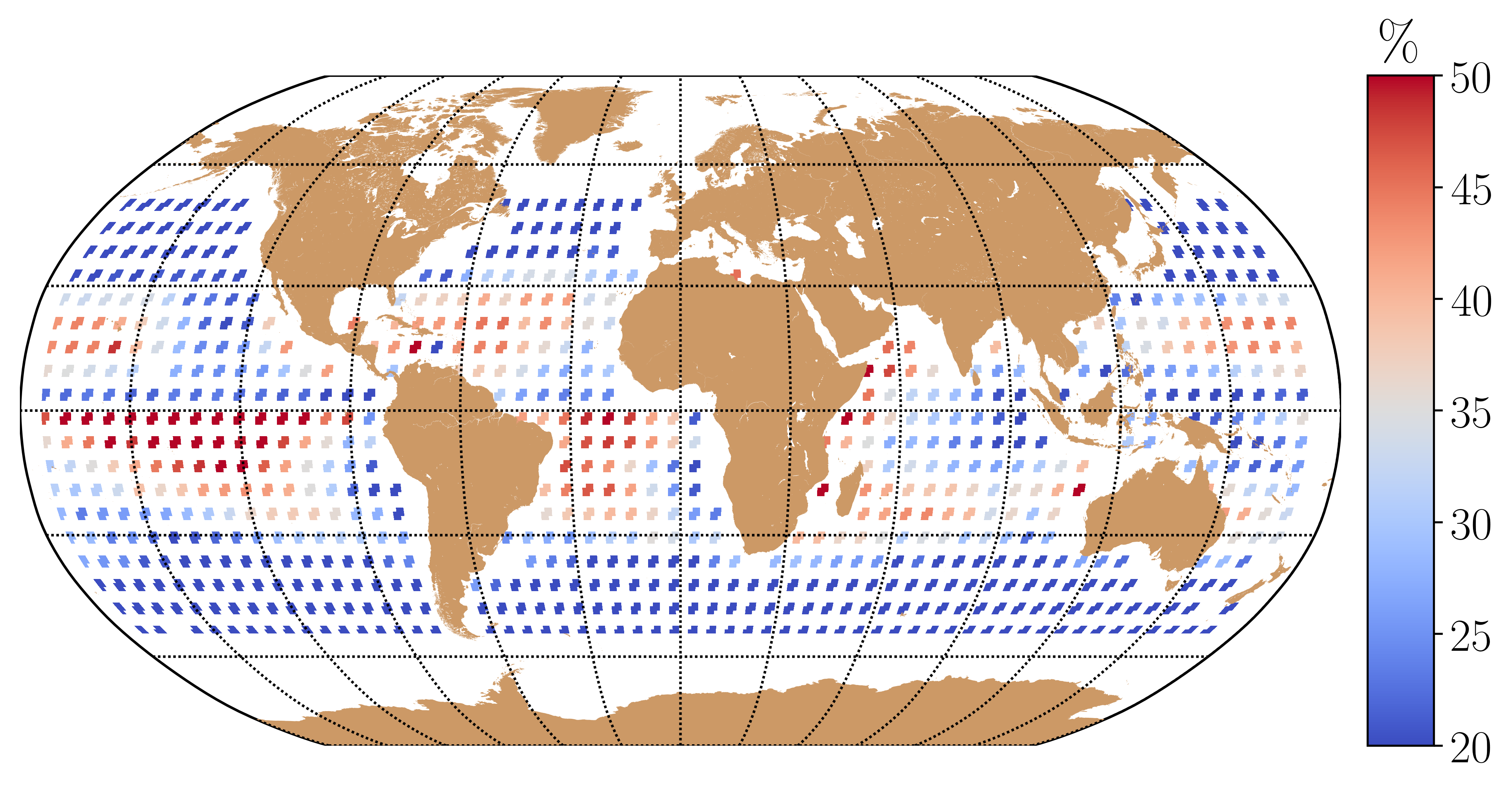}
\caption{Percentage of the time (number of days over the entire 2002-2019 study period) that the MODIS Aqua sensor reported SST estimates for all global points considered in Figure \ref{fig:globalskill}.  }
\label{fig:perccove}
\end{figure}

Figure~\ref{fig:globalskill} indicates some spatial variations in performance. In general MAE is lower in the inter-tropics region than in southern or northern latitudes. This effect is more pronounced when we consider MAPE values due to lower ambient temperatures making relative differences more pronounced. Further analysis indicates that this spatial bias is largely driven by reduced data availability in locations away from the tropics. Figure \ref{fig:perccove} presents the percentage of days for which data was available for the study period (100\% indicates that data is available every day). We see a distinct pattern of higher data availability over the inter-tropical regions which is possibly a result of increased cloud coverage in southern and northern latitudes \citep{ruiz2013assessment, o2019observational}. For all locations we replaced missing data using linear interpolation which enables the models to act on data-sparse regions but limits the amount of true data available to learn the complex SST relationship. Further all error metrics were computed on the raw data (without linear interpolation) which biases the evaluation further towards locations with higher data coverage.


\subsection{Discussion of results}
\label{sec:r&d}
Time-series forecasts are vital in many areas of scientific, industrial, and economic 
activity. Many ML methods have been applied to such problems and the advantages of RNN-type approaches are well documented. The ability of these DL algorithms to implicitly 
include effects from preceding time steps is intuitively a natural fit. However, this study 
demonstrated that when the relationship between predicted values was not based solely on AR features -- well-designed feature selection in conjunction with simpler ML methods capable to more rapidly adjust to short-scale fluctuations outperformed the DL approaches.

Table~\ref{tab:model_performance} demonstrated that the important features needed to
predict at time $t+1$ are SST values at time $t$ (and values at earlier 
time steps dependent on selected AR features), atmospheric information
at time $t+1$ (and potentially AR features of those), and time of year information. Hence, prediction 
required inclusion of AR features while inferring relationships 
between forecasted values of atmospheric data and the response variable, SST. Figure~\ref{fig:different_models} illustrated that the LSTM model failed to adequately learn the 
relationship between explanatory variables and SST. Specifically, the model 
closely approximated seasonal behaviour (i.\,e.,~the long-term characteristics of the SST) while failing to capture high-frequency variations (i.\,e.,~variations in response 
to atmospheric inputs). In effect, the DL approach maintained ``memory'' of the 
long-term SST trends to the detriment of incorporating effects of shorter 
time scales. A~more focused feature-engineering module that guided the data-length 
fed to the LSTM model may improve performance. However, this contravenes 
the philosophy of RNN-type approaches that aims to implicitly learn the nature of cyclic data. 
Another point worth noting is that DL approaches have a larger 
appetite for training data than some of the simpler models adopted. Some 
reduction in MAPE may be possible by extending the size of the training data. Again, however, 
when evaluating different modelling approaches, aspects such as computational complexity 
and ability to learn on smaller datasets are key points that demand consideration
 (further, there are practical limits on amount of available data). 

This study considered a framework to develop a transportable model suite applied to a nonlinear, real-world dataset. Key points considered were design of an automatic feature-engineering module, which, together with a standard hyperparameter optimisation routine, 
facilitated ready deployment at disparate geographical locations. Results demonstrated that 
the different models adopted had inherent characteristics that governed accuracy and level of regularisation or overfit to training data.

\begin{figure}[h!]
\centering
\includegraphics[width=\textwidth]{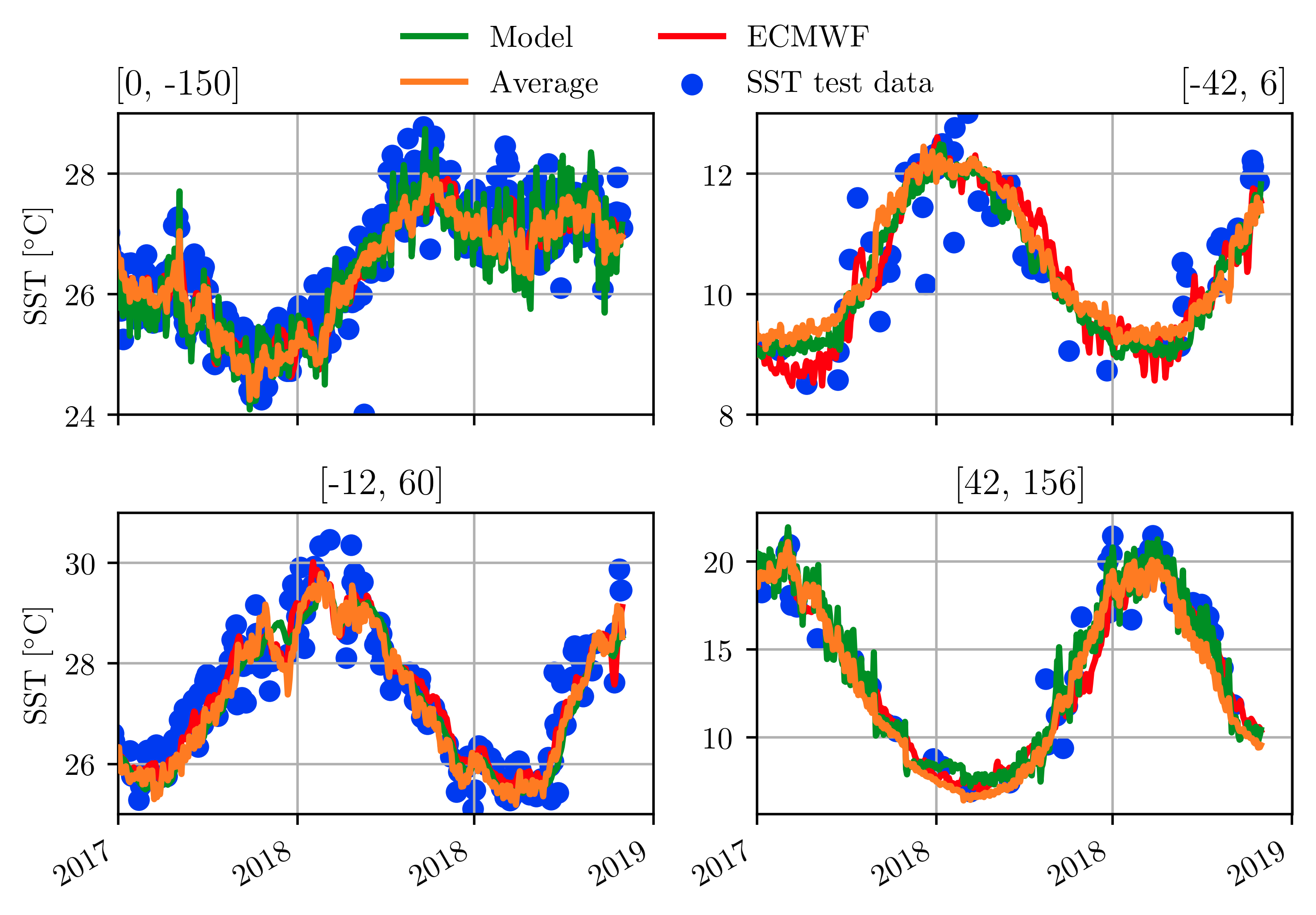}
\caption{Time series plots comparing the performance of individual model (green line) and weighted ensemble average of  GAM, RF, XGBoost, and MLP models (orange line) against 1) ECMWF model estimate at nearest grid cell (red line) and 2) SST test data from MODIS satellite (blue circles). The location of the four points considered are illustrated in Figure 
\ref{fig:globalskill} and described in more detail in Table \ref{tab:ts_mse} (note that subfigure title denote the latitude and longitude coordinates). For each of the four plots the orange line presents a different model to provide an illustration of different model characteristics, namely (from top-left) GAM, RF, XGBoost and MLP. }
\label{fig:simulation_comparison}
\end{figure}

We compared performance of ML models with a state-of-the-art 
physics-based approach from ECMWF. As expected, the physics-based model provided close agreement with satellite measurements -- the ECMWF prediction is a reanalysis product 
which assimilates measurement (including satellite) data
daily to update the accuracy of the product.
This study demonstrated, however, that the machine learning based approaches achieve accuracy comparable to ECMWF model, at a fraction of the computational expense.
Aggregating the models improved the robustness of this approach and served to regularise small-scale fluctuations or seasonal biases in  individual models. 
Figure~\ref{fig:simulation_comparison} compares the ensemble predictions to the ECMWF results,  satellite measured SST and predictions from selected ML model at four locations across the globe (location details provided in Figure \ref{fig:globalskill} and Table \ref{tab:ts_mse}). The four plots illustrate the varying temporal characteristics of SST data at different geographical points and the performance of ML models to capture those characteristics. Generally the models are seen to capture both the seasonal patterns and shorter-scale fluctuations (e.g. unseasonably warm autumn temperatures at location [0, -150]). Individual model prediction (green line) provides good predictive skill comparable to ECMWF, while the aggregated model is `smoother' (possibly more robust to short-scale fluctuations), while achieving comparable accuracy.

\begin{table}[h!]
  \caption{MAE (top table) and MAPE (bottom table)  for individual models and ensemble average of all models (GAM, RF, XGBoost, and MLP) for  the selected number of locations presented in Figure \ref{fig:simulation_comparison} . Forecast skill of ECMWF model also presented for illustrative purposes.  }
  \centering
  \begin{tabular}{c c c c c c c c }
  \toprule
	\textbf{Lat} & \textbf{Lon} & \textbf{GAM} & \textbf{RF} & \textbf{XGBoost} & \textbf{MLP} & \textbf{Ens. Ave.} & \textbf{ECMWF} \\ \midrule
	0 & -150  & 0.41 & 0.44 & 0.43 & 0.42 & 0.38 & 0.33 \\ 
      -42 &     6  & 0.6 & 0.73 &0.88 & 0.7 & 0.67 & 0.63  \\
      -12 &   60  & 0.48 & 0.49 & 0.55 & 0.50 & 0.45 & 0.37  \\ 
       42 & 156  &  1.96 & 1.31 & 1.53 & 1.09 & 1.16 & 1.03  \\  
\midrule
 & & &  & \multicolumn{1}{c}{\textbf{MAPE} } & &  &\\  \midrule

   	0 & -150  & 1.52 & 1.64 & 1.61 & 1.58 & 1.42  &  1.27 \\ 
      -42 &     6  &  5.38 & 6.80 & 8.0 &   6.52 & 6.16  & 5.78 \\
      -12 &   60  &  1.77 & 1.79 & 2.01 & 1.86 & 1.66  & 1.38 \\ 
       42 & 156  &  12.5 & 8.27 & 10.4 & 7.26 & 7.56 &  6.62  \\  
 \bottomrule
\end{tabular}
\label{tab:ts_mse}
\end{table}

Classical works on ensemble forecasting demonstrated that the ensemble mean should give a better forecast than a single deterministic forecast  \citep{epstein1969stochastic, leith1974theoretical}. 
Assigning inverse MAPE weights to individual models provides a simple and effective method to rank model contributions based on performance. 
To illustrate forecast skill of different models, Table \ref{tab:ts_mse} presents MAE and MAPE for each individual model, an ensemble model aggregation, and the ECMWF estimate against MODIS measurements for the locations plotted in Figure \ref{fig:simulation_comparison}. Results demonstrate that the variation in error of individual models can be "regularised" by the ensemble approach. We observe that individual models perform better at different locations (e.g. GAM performs best at location [-42,6], while MLP performs best at [42, 156]), illustrating the concept of the "no free lunch" - no single machine learning algorithm necessarily outperforms all others and one must select the appropriate algorithm for the problem. However, the ensemble averaging approach outperforms individual models providing a framework to improve average predictive skill. 
The low computational cost of prediction enabled by machine learning is particularly amenable to ensemble modelling approaches where multiple models can be readily deployed \citep{odonncha2018integrated,o2019ensemble}.

Interrogating temporal evolution of model error over the 18-month test period demonstrated some biases in individual models – e.g. GAM outperformed RF during the summer period but is significantly poorer during periods of lower temperature. The ensemble aggregation framework we implemented reduced error over the duration of the test period compared to arbitrarily selected individual models, but more importantly, also served to reduce error and biases at distinct periods of the prediction window. Table \ref{tab:seas_mse} presents seasonal MSE against satellite data for each individual model and the ensemble aggregation over the duration of the study period averaged over the same locations as Figure~\ref{fig:simulation_comparison}. 

\begin{table}[h!]
  \caption{Seasonal MSE for individual models and ensemble average}
  \centering
  \begin{tabular}{c c c c c c}
  \toprule
	\textbf{Season} & \textbf{GAM} & \textbf{RF} & \textbf{XGBoost} & \textbf{MLP} & \textbf{Ens. Ave.}\\ \midrule
	Spring & 1.37  & 1.04 & 1.01 & 0.90 & 0.94 \\ 
    Summer & 0.18  & 0.40 & 0.45 & 0.68 & 0.30 \\
	Autumn & 0.06  & 0.10 & 0.40 & 0.80 & 0.17 \\ 
	Winter & 0.26  & 0.14 & 0.39 & 0.34 & 0.22 \\  
\bottomrule
  \end{tabular}
\label{tab:seas_mse}
\end{table}


This study presented a time-series forecasting framework applied to satellite measurement of SST.  We considered the SST data as a set of disparate points. In 
reality, the ocean surface more closely resembles an image with interaction between neighbouring
points. Results demonstrated that treating the data as 
distinct time-series points provided good results. However, scope exists to combine this approach 
with image-processing techniques such as convolutional neural networks (CNNs) to incorporate neighbouring effects into predictions. Future work will explore the viability and value of combining CNNs with time-series forecasting models to further improve the robustness of the framework.

\section{Conclusions}
\label{sec:concl}
This paper demonstrates the viability of applying ML based approaches, addressing transportability, biases and robustness by combining feature selection and disparate models with specific characteristics in a weighted aggregation based on average model performance. 
This study aimed to assess the ability of data-driven approaches to accurately predict SST – characterised by seasonal patterns, temporal dependencies and short-term fluctuations. 
Results demonstrate comparable performance to physics-based model simulations with low computational cost, and which is easily parametrised to other geographical locations.  
The low computational cost of the approach has many advantages. First, it enables 
separation of SST forecasting models from HPC centres -- the suite of models presented here 
can be trained on a laptop and applied to any geographic location. 
Once trained, the inference step is of negligible computational expense and can be readily 
deployed on edge-type devices (e.g. in-situ devices deployed in the ocean).
Deploying large-scale models is a complex task highly dependent on user skill
to correctly configure and parametrise to specific locations. Data-driven approaches can present
an alternative approach that enables rapid prediction, contingent on availability of sufficient data. 

\section*{Acknowledgements}
\noindent
This project has received funding from the European Union’s Horizon 2020
research and innovation programme as part of the RIA GAIN project under grant
agreement No. 773330.

\clearpage

\end{document}